# A press-to-contact acoustic sensor for measuring elasticity and ageing of extracellular matrix

Jenny Overton[1], Arthur Kissin[1], William Bains, Adrian Stevenson[*]

*Department of Chemical Engineering and Biotechnology, University of Cambridge*

**Abstract**

We report an acoustic approach to measure the elastic properties of an extracellular matrix (ECM). This approach uses a short range evanescent waves to inject sound into the test material over a short distance using an electroded quartz crystal, also known as a quartz crystal microbalance (QCM). These measurements are applied with a press-to-contact approach to determine the elastic properties reliably and rapidly. The elasticity of ECM is connected to material chemistry which is linked to the tissue type and history. Measuring the minimum relative amplitude of the signal from the QCM shows that the acoustic response is strongly correlated to tissue type and age. *Keywords:* ECM, QCM, tissue elasticity, ageing

## 1. Introduction

Elasticity measurements for bio materials such as extracellular matrix (ECM) need a markedly different measurement strategy relative to the predictable behaviour of inorganic fluids, metals and ceramics which usually have known and uniform properties. Issues with bio material and biofilm measurements have been evident since the 1970s when tracking antibody antigen binding reactions with acoustic waves became common (Shons et al. (1972)). Also sample shape remains a constraint with MEMS to measure hydrogel films (Corbin et al. (2013) ) or acoustic imaging systems to cell cultures (Kuo et al. (2017)), both preferring samples in the form of defined sheets.

[*]Correspondance to: acs14@cam.ac.uk
[1]Both authors contributed equally

The ExtraCellular Matrix (ECM) is central to the mechanical properties of the tissues and is therefore important in a range of biological and medical studies. It comprises of a wide range of long chain proteins including collagen and elastin which accumulate cross-links over time (Wang and Thampatty (2006)). These cross-links occur through glycation and enzyme activity and cause the ECM to be less flexible and open to remodelling (Robins (2007)).

Degradation and cross-linking of the ECM in old age in humans is known to be linked to ageing related illnesses such as cardiovascular disease and Alzheimer's. The main method of characterisation of ECM at present is chemical (Cox and Erler (2011)) - the chemistry of the ECM has been well documented in the literature and the cross-linking in a sample can be chemically defined but this requires destructive techniques which currently require full sample break down (due to a lack of effective imaging agents). Using immunohistochemistry or immunofluorescence to stain tissues for biochemical markers is a common method to establish changes in the ECM composition, including the occurrence of some cross-links.

An alternative to chemical characterisation is measurement of the mechanical properties of the ECM. The ECM's mechanical properties directly relate to bulk flexibility and resilance of the tissue, but also correlate with the abilities of cells within the ECM to grow, function appropriately, and remodel the ECM itself. Recently, atomic force microscopy (AFM) has been used to obtain high-resolution images of ECM structure (Graham et al. (2010)) which allow the ECM elasticity and stiffness to be evaluated on the micron scale. However, this is not suitable for bulk measurements which are crucial in a number of cases. Other techniques such as shear rheology are commonly used to measure ECM stiffness at a macroscopic level but the process is destructive, and therefore not always desirable.

A good starting point for macro-scale mechanical measurement of ECM builds on lower frequency travelling wave and pulsed wave acoustic methods for characterising material samples. If a sample is placed in the path of these waves the hydrophone will detect a measurable change. However the dimensions, surfaces and sample volume must be known in order to calculate the mechanical properties of the sample, so for the typical variable and irregular tissue ECM samples this familiar acoustic approach is not viable. In this paper we present a press-to-contact shear wave geometry which uses a novel configuration of a facile, compact and low cost piezoelectric crystal to garner elasticity information, without unwanted dimensional and reflectivity problems.

Low power acoustic techniques have been adapted to study the intrinsic properties of human biological materials such as aortic tissue (Akhtar et al. (2008)). In particular it allows the ECM's properties to be tested without destroying structure or breaking up the sample. Scanning acoustic microscopy (SAM) has been used to study soft biological tissues (Zhao et al. (2011)), but it has numerous issues including high costs, long setup times, noise from wave reflections, an insensitivity to macro-scale properties and the need for the specimen to be immersed in water (Doroski et al. (2007)). There is a need for an alternative sensing geometry that can deploy the advantages of acoustics routinely alongside chemical analysis to expose the ECM's mechanical properties and its role. This aim of this work is to address this gap and adapt a routinely available acoustic device to probe the mechanical properties of the ECM.

## 2. Theory

The QCM is the most common commercialized acoustic sensor and is traditionally used to investigate nanometre thick films to determine their mass and viscoelastic properties. It is also used as a frequency control device as the resonance frequency can be adjusted very precisely. The crystal typically operates in a Thickness Shear Mode (TSM). Two electrodes coat the upper and lower surfaces of a quartz disc and when an alternating voltage is applied to the electrodes, the shear resonance of the disc is stimulated.

The response of QCM sensors to a change in the external environment involves a shift in the resonance frequency and an amplitude change of the resonance peak. It is this shift in the resonance frequency peak which is most commonly measure in acoustic sensors (Hu et al. (2012)).

The most recognized mathematical model for the QCM is the Sauerbrey equation (Sauerbrey (1959)):

$$\frac{\Delta f}{f_0} = \frac{\Delta d_l}{d_Q} = \frac{\Delta m_l}{\rho_Q A d_Q} \qquad (1)$$

where $f$ is the frequency shift, $f_0$ is the operating frequency, $d_l$ and $m_l$ are the thickness and mass of the loaded film, $d_Q$, $\rho_Q$ and $A$ are the thickness, density and area of the quartz disc respectively. This model assumes a very thin rigid film on the top of the quartz disc, and was improved by Kanazawa and Gordon (Kanazawa and Gordon (1985)) to allow for the material detected to be in an ideal Newtonian fluid:

$$\Delta f = f_0^{3/2} \frac{\eta \rho}{\pi \mu_0 \rho_Q} \qquad (2)$$

Where η and ρ are the viscosity and density of the Newtonian fluid and $\mu_0$ and $\rho_Q$ are the elasticity and density of the quartz respectively. Many models suggested divide the system into a mechanical and electric circuit model (Hu et al. (2012)). Figure 1A shows the Butterworth Van Dyke (BVD) equivalent circuit impedance model of a TSM resonator. The advantage of BVD is that the behaviour of the resonator is generally probed using electrical signals from an impedance analyser.

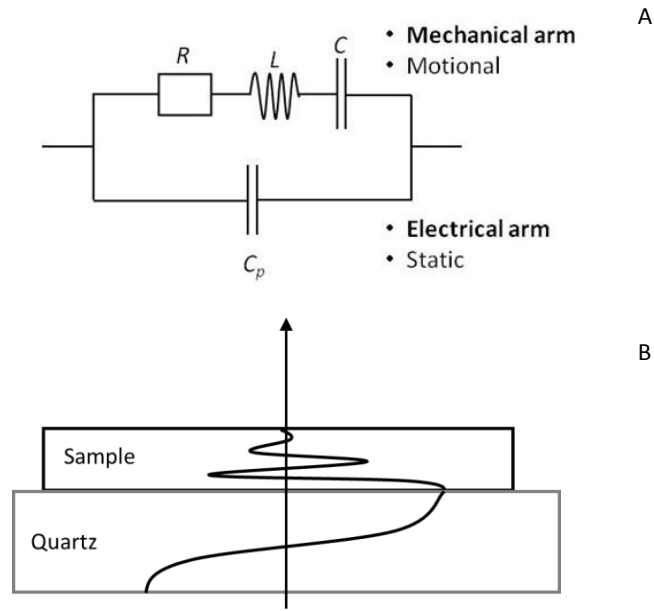

Figure 1: A: BVD equivalent circuit impedance model of a TSM resonator adapted from Kanazawa and Cho (2009). B: The target geometry showing the ECM sample contacting the TSM device and the ECM uptaking a shear wave which due to damping from scattering and viscous interactions does not penetrate through the whole sample, adding immunity from reflection

The parallel capacitance $C_p$, is the constant capacitance which reflects the two electrodes on the crystal surface as well as any external capacitances which occur due to connecting the QCM to the instrumentation. The motional arm represents the mechanical losses of the resonant system (Kanazawa and Cho (2009)). It includes the inductance $L$, a series capacitance $C$, and a series resistance $R$. These represent the mass, elasticity and viscosity of the quartz disc respectively (Hu et al. (2012)). The mechanical arm is altered when a sample is loaded onto the quartz disc which is sometimes represented as an additional motional inductor and resistor on the mechanical arm to represent the mass and viscoelastic material properties. The surface-loaded resonator has an impedance of:

$$Z_m = (R + \delta R) + j\omega(L + \delta L) + \frac{1}{j\omega C} \quad (3)$$

where $Z_m$ is the motional impedance, and R and L are the real motional resistance and inductance due to the sample on the surface. It can be seen that the addition of a sample changes R which is directly related to the mechanical properties of the sample such as the fluid viscosity. The critical feature of this system is the propagation of an attenuated shear wave through the contacting material. As long as there is sufficient attenuation of the sound wave in the sample at the ultrasonic frequency of the crystal, the wave will decay within the medium. It will not traverse the whole material and reflect off a boundary and produce deleterious noise. Unlike longitudinal waves, shear waves have this important noise reduction property. Hence the combination of shear waves of a specific frequency which decays within the material is desirable (see Figure 1B). Assuming contact pressure between the crystal and the sample can be managed, then the net result is that the acoustic wave and the electrical properties of the QCM probe is unambiguously linked to the sample properties.

### 3. Materials and Methods

*3.1. QCM probes*

Quartz crystals were purchased from Maplin Electronics UK. When supplied, the crystals are in protective metal cases which were removed to access the crystal surfaces. To increase the mechanical stability of the crystal, the base of the protective case was left attached to the connecting wires as seen in Figure 2C . Crystals with resonant frequencies of 5-12 MHz were tested to determine the optimum crystal frequency. The 10 MHz crystals was seen to have the most reliable response and was selected for further experiments
(See Supplementary Information)

*3.2. Polyacrylamide analogues*

ECM is decellularised tissue and resembles the structure of a polymer gel rather than tissue. A polyacrylamide gel was chosen as defined, sterile polymer analogue of the ECM because of its highly hydrated, hydrophilic, cross-linked polymer network forming nature. It has been used previously alongside ECM for example to measure intercellular tensional forces (Tseng et al. (2012)) or matrix stiffness where a thin layer of ECM matrix is overlaid on polyacrylamide (Cox and Erler (2011)).

The mechanical properties of polyacrylamide gel can be altered by controlling the acrylamide monomer/BIS cross linker ratio to form gels with different degrees of cross-linking (Wong et al. (2003)). A range of polyacrylamide gels with varying levels of cross-linking were used to create a data set with which to compare the ECM (Bio-Rad protocol used for making gels).

The polyacrylamide gels produced for initial tests used 50% acrylamide/BIS (29:1) (50 g of acrylamide/BIS in a 29:1 ratio dissolved in 100 ml of sterile water). The T value (which defines the solidity of the gel through the concentration of polymer in the solution) was varied between 6% and 40% by adding varying volumes to $25\mu l$ of freshly made catalyst (100 mg ammonium persulfate in 1 ml of water) and $5\mu l$ of TEMED (N, N, N tetramethylethylenediamine). The solution was made up to 5ml using sterile water. The TEMED accelerates the rate of formation of free radicals from persulfate and these in turn catalyse polymerisation (Tanaka (1979)). For each percentage value, two 2 ml samples were produced and left to set at room temperature for three days.

### 3.3. Tissue samples

Tissue samples of young chicken were obtained from UK supermarkets. These were estimated to be between 35 and 63 days old based on available standards for UK supermarkets (Ba´eza et al. (2012)). Older chicken samples came from a pet chicken estimated to be around the age of two at the time of death. Canine samples were donated by the University of Cambridge Veterinary School. Each tissue sample was decellularised using the protocol outlined in (Mirzarafie et al. (2014)) to obtain pure ECM.

### 3.4. Instrumentation
### 3.4.1. Resonance frequency measurements

The effect of a sample on crystal resonance is typically measured using an impedance analyser, which inputs an electrical frequency through the electrodes. An Agilent Impedance Analyzer 4291B was used for this work.

When the input electrical frequency is that of the resonance frequency of the crystal, substantially greater energy is stored so the impedance of the circuit is at a maximum at this point. A contacting sample removes energy from the crystal, changing the shape and position of the resonance according to its mechanical properties. The impedance analyser is able to sweep over a range of frequencies, measuring the relative amplitude (impedance) of the wave at each frequency and plotting a trace to show clearly the systems response of the measured frequencies. By performing this analysis both with and without the sample, the change in peak frequency and shape can be measured and analysed (Kanazawa and Cho (2009)). For comparison purposes, impedance is labelled as relative amplitude in the figures.

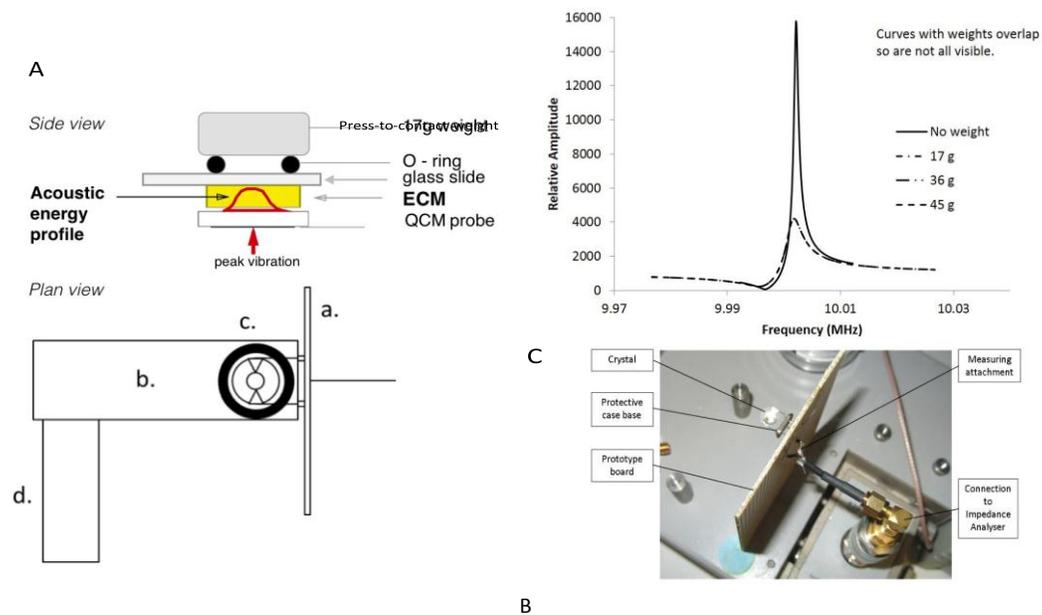

Figure 2: A: Diagrammatic representation of the press-to-contact arrangement for the ECM sample on the QCM device. The plan view shows (a) the crystal platform, soldered to the prototyping board connected to the impedance analyser (b) Glass slide placed between clamp support stand and sample on crystal surface - the smooth flat surface ensures the radially applied load presses evenly on the sample while also allowing the sample to be observed throughout. (c) cylindrical weight which allows force to be applied radially around the sample to prevent point loading. (d) clamp stand support B: Frequency recorded when different weights were applied to 20% 29:1 polyacrylamide gel on a 10 MHz crystal. All three curves for the applied weights overlap entirely. C: Photograph of the test geometry (without sample and press-to-contact weight)

*3.4.2. Press-to-contact QCM (PCQCM) device configuration*

The link between the sample and QCM probe is critical to the reproducibility of the results, so the following configuration was adapted to achieve reproducibility and stability. This approach made use of the technical note by (Kanazawa and Cho (2009)) that the crystal resonance is not expected to be uniform, having a peak displacement in the middle and falling away at the edges (See Fig 2A, in red). Therefore, samples are targeted to cover the electrode completely.

In order to investigate the effect of contact force on repeatability, polyacrylamide gel was used for validating the QCM probe. With uniform force application around the device edges via a o-ring, it was found a loading weight of 17g and above gave a reproducible response (see Fig 2B).

Key wiring considerations were the distance between the crystal and the measuring attachment, which causes slight changes in the measured frequency. Also, the crystals are very delicate and susceptible to damage. Therefore a rig was prepared to attach the crystal in a regular and stable manner, as shown in Fig 2.

The final probe comprised a QCM device connected to an impedance analyser and a weight coupled to the sample so that constant pressure is applied at the contact point between the sample and the crystal. The collected data involved the tracking the change in resonance frequency and amplitude associated with sample contact. As an additional support structure was used, there were no further mechanical shifts that could move or affect the radio-frequency wiring, which is important for data reproducibility.

*3.5. Data collection and Minimum Relative Amplitude (MRA) analysis*

The most common method of data analysis for QCM acoustic sensors is to look at the peak frequency shift as described in the previous sections (Kanazawa and Cho (2009)). Different samples gave very different frequencyamplitude relationships when placed on the crystal as shown in Figure 3B. The clean crystal gives a clearly defined peak (shown in the figure as a vertical line for clarity), as did the polyacrylamide gel samples. However, the skin ECM samples gave traces with almost equal peak and trough heights and some muscle samples (such as the chicken muscle shown here) gave a defined trough but no distinguishable peak. Initially, it was thought that peak values should be taken for all samples but this was not possible as often with ECM samples the peak shifted to the right of the clean crystal resonance frequency, which is not physically interpretable.

The difference in relationship could be due to interference from the electrical arm masking the mechanical properties of the sample. However the measurement does provide a parallel frequency peak alongside the trough (Wohltjen et al. (1997)). Although this relative amplitude is smaller, it is directly correlated to the pure mechanical resonance properties of the sample according to the equivalent circuit model. This Minimum Relative

Amplitude (MRA) metric 'allowed for all sample types to be mechanically tested by the PCQCM probe. All subsequent experiments using ECM samples follow this MRA trough analysis.

Figure 3A shows a clear relation between different amounts of crosslinking and the acoustic response. This suggests the press to contact QCM probe is responding to mechanical property change induced by cross-linking and is therefore suitably configured for the relative testing of ECM samples from chicken and canine muscle and skin.

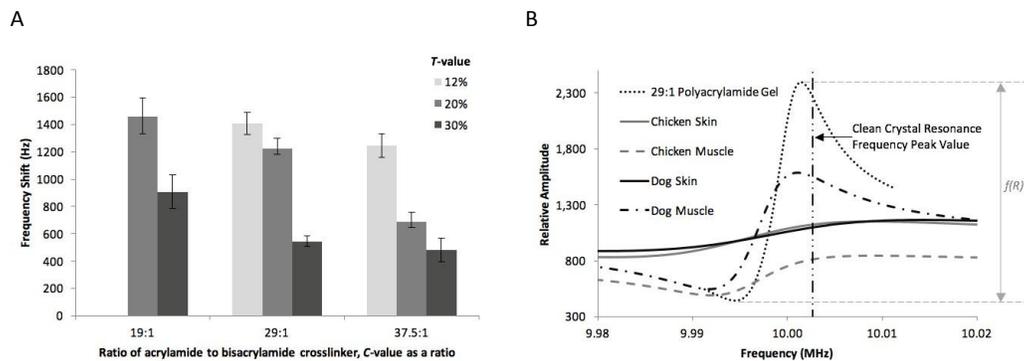

Figure 3: A: Results from testing a sensor with polyacrylamide gel show that it can distinguish between both varying C-value and T-value, both of which are indicators of mechanical properties. It shows an increase in frequency shift as stiffness is increased by increasing the amount of cross-linking at a higher concentration. B: Raw data output for different sample types to highlight the differences in the frequency-amplitude relationship

## 4. Results

### 4.1. PCQCM response to ECM samples

#### 4.1.1. Tissue type behaviour

As discussed in Section 3.5, for each sample tested the minimum relative amplitude (MRA) was plotted against its corresponding frequency as shown in Figure 4A-D . A correlation between the amplitude and frequency at this minimum relative amplitude holds for many different samples. We propose that as a sample stiffens and cross-linking is increased, it would move up and along the best fit line of the graph to the left. The average MRA and associated frequency was plotted for each sample type in Figure 4. This shows that the devised PCQCM apparatus and method of data analysis can clearly distinguish different types of samples as required.

*4.1.2. Tissue age behaviour*

Old and young chicken samples were analyzed (see Fig 4) . All samples were frozen and thawed prior to the decellurisation protocol. This data indicates that the PCQCM probe can show differences between chicken samples of varying ages. However, it cannot yet be proven conclusively that this is due to differences in ECM cross-linking.

*4.1.3. Tissue freeze-thaw shifts*

The effect of freeze-thaw cycles on the mechanical properties of the samples is shown in Fig 4. Chicken samples were tested both where the tissue had been frozen then thawed prior to decellularisation and where the tissue was decellularised from fresh. Further samples of chicken and canine skin and muscle were also tested. These samples had been frozen and thawed prior to decellularised. The data labeled "refreeze" was put through an additional freeze-thaw cycle. The arrows indicate the movement of the sample after the refreezing process. The data also shows that there are differences associated with freezing prior to, and post, the decellularisation protocol.

## 5. Discussion

*5.1. The acoustic response of the PCQCM device to the elastic change*

The principle focus of this work is configuring an acoustic probe to recover important mechanical information from the extracellular matrix. The underlying principle of an acoustic measurement is summarised as an elastic spring that connects two regions together. As the acoustic wave passes its energy through the spring, the equivalent spring constant of the material defines its response - the stiffer the spring the greater the chemical interaction forces. Increasing elasticity can be seen as greater wave speed and less damping, which means crosslinking assists the passage of the wave. The wave polarisation is also critical. For a wave polarised in the compressional direction the wave motion is forward and back along the propagation path. On the other

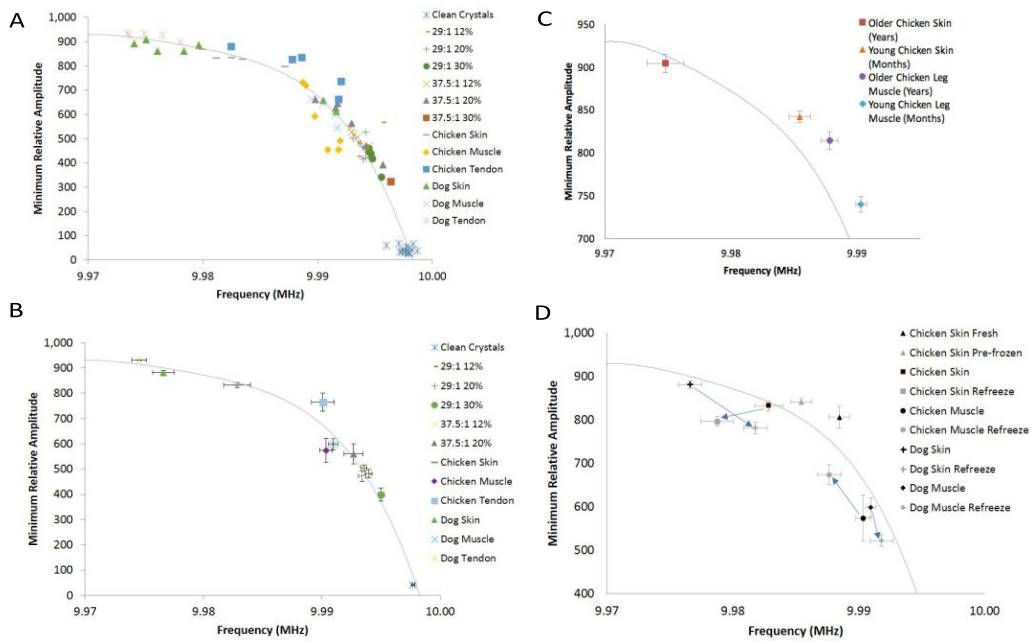

Figure 4: Minimum Relative Amplitude measured (see section 3.5) vs Frequency. The line shown in each case is the line of best fit, and error bars show the standard error of the mean. A: Individual data points for different samples show a trendline associated with acoustic energy loss from the QCM. B: The average values for each sample type show high reproducibility and the utility of PCQCM for performing easily accessible laboratory measurements of mechanical properties. C: Comparison of skin and leg muscle samples from old chicken (around 2yrs) and young chicken (around 1.5mths). In each case the older sample shifts the resonance frequency to the left D: Comparison of a range of chicken and canine ECM samples. Arrows show direction of sample movement after a freeze-thaw cycle

hand, a shear wave motion is side to side relative to the propagation path, which is appealing as it significantly reduces the distances travelled lowering acoustic noise.

We have described a press-to-contact QCM (PCQCM) for measurement of ECM elasticity. The profile of the acoustic field from the device is an important design aspect. It has a strong central vibration and minimal edge movement, so a weight can be applied to the edges via an o-ring without curtailing motion. The central press to contact force then provides wave transmission into the ECM so it can interact with crosslinked regions and report back via changes in the resonance curve.

Normally a QCM wave decays only a micron into the fluid, but next to a stiff material such as the ECM the wave connects strongly through the neighbouring region of the material. However it was not possible to quantify this wave penetration. We estimate it exceeds a micron due to its significant stiffness, but is less than the sample size as no internal reflections from nearby boundaries were observed - around $1mm$ of propagation into the ECM is likely.

*5.2. The structure and dynamics of the extracellular matrix*

At this point all evidence supported a link between PCQCM behaviour and sample crosslinking. This was consistent with expectations set from the acoustic wave physics and from the data collected from the polyacrylamide gels. However insufficient mechanical data on the samples themselves prevented detailed numerical comparisons, as their irregular shape was challenging to characterise with other techniques.

Figure 4 indicates that a correlation between the frequency shift and the minimum relative amplitude exists. Energy transfer out of the QCM is a likely explanation for this curve. The consistent reduction in frequency of the crystal is associated with an energy loss. This is undoubtedly associated with the absorption of that energy by the contacting ECM. Clear differences between hydrogels, skin, muscle and tendon are seen, which is consistent with variable leakage of wave energy derived from mechanical differences.

Older ECM sample generate larger frequency and amplitude shifts. This is explained by energy leakage stimulating greater amplitude and frequency shifts. The energy transfer model explains this relationship. Here stiffness of the ECM relative to the extremely stiff crystal modulates this energy leakage. The greater the elasticity difference, the lower the energy leakage, hence the form of this curve can be explained by an impedance transfer model. Figure 4D shows that the ageing effect is the outstanding feature amongst several other situational variables, confirming the excellent specificity of the PCQCM approach.

## 6. Conclusions

This work shows that the mechanical properties of a sample can be investigated using acoustic measurement. The press-to-contact QCM probe working in thickness shear mode (TSM) at a defined contact pressure and attached to an impedance analyzer, is able to clearly distinguish samples of different types in a consistent fashion, and overcomes the typical challenges associated with measuring mechanical properties of organic materials and biofilms.

Variabilities of the wetted contact area has already been minimized by the application of a constant force. However, the system could still be improved further, by using a microcontroller such as a Raspberry Pi could be configured to track sample crosslinking on crystal surface as a frequency-amplitude ratio measured over time. The collection of more data under controlled conditions can be used to establish a formulaic linkage between the recorded frequency shift of the minimum amplitude and cross-linking process. The system can also be integrated substantially by using an FPGA network analyser which is a fraction of the size and weight of the impedance analyzer used here. Use of thicker crystals, and hydrophobic coatings that prevent accumulation and fouling of the surface would also improve the system, by improving robustness and reducing noise.

This PCQCM probe is able to show a correlation between mechanical properties, the minimum relative amplitude and the associated frequency. Chicken samples with an older age at time of death have lower frequencies (and higher relative amplitudes) than younger chickens as expected, which supports increased cross-linking hypothesis. The effect of freezing samples both before and after the decellularisation protocol has a marked effect however further experiments are required to determine the consequences of this. Similarly for polyacrylamide gels it is observed that the peak amplitude frequency shift can give a reasonable measurement of relative cross-linking, but that polyacrylamide overall is not a good polymer model for comparison with ECM cross-linking as it is closer to a liquid system.

This acoustic method shows significant promise for estimating the relative mechanical properties of decellularised extracellular matrix samples. Data shows a consistent increased stiffness in older chicken samples and no issues or complication were found due to noise from boundaries, which this shear wave probe avoids. Provided the contact forces can be maintained, this is a simple approach that is amenable to portable device testing of skin. With further adaptation it has the potential for tracking the elasticity of the skin in the context of animal and human studies involving live subjects.

The authors would like to thank Dr Isik Ustok and the SENS Research Foundation for assisting with this study.